\documentclass[%
 aip,
 amsmath,amssymb,
 reprint,%
]{revtex4-1}

\usepackage{lipsum}
\usepackage{mwe}

\usepackage{graphicx}
\usepackage{dcolumn}
\usepackage{bm}

\usepackage[utf8]{inputenc}
\usepackage[T1]{fontenc}
\usepackage{mathptmx}
\usepackage{etoolbox}
\usepackage{subcaption}
\usepackage{float}

\makeatletter
\def\@email#1#2{%
 \endgroup
 \patchcmd{\titleblock@produce}
  {\frontmatter@RRAPformat}
  {\frontmatter@RRAPformat{\produce@RRAP{*#1\href{mailto:#2}{#2}}}\frontmatter@RRAPformat}
  {}{}
}%
\makeatother

\newcommand{\beq}{\begin{eqnarray}}
\newcommand{\eeq}{\end{eqnarray}}

\newcommand{\vp}{{v_\perp}}

\newcommand{\B}{{\bf B}}
\newcommand{\bb}{{\bf b}}

\newcommand{\tx}{{\tilde{x}}}
\newcommand{\vz}{v_{\parallel}}
\newcommand{\fd}[2]{\frac{\displaystyle #1}{\displaystyle #2}}

\newcommand{\avep}[1]{\left< #1 \right>_{\phi}}

\def\eq#1{(\ref{eq:#1})}
\def\ddt#1#2{\partial #1 /  \partial #2}
\def\dd#1#2{\fd{\partial #1}{  \partial #2}}

\begin{document}

\preprint{AIP/123-QED}

\title{ Pitch Angle Scattering of Fast Particles by Low Frequency Magnetic Fluctuations}
\author{Yi Xu}
\author{Jan Egedal}%
 \email{egedal@wisc.edu}
\affiliation{ 
$^1$Department of Physics, University of Wisconsin - Madison, 1150 University Ave, Madison, WI, 53706, USA.
}%

\date{\today}

\begin{abstract}
The adiabatic invariance  of the magnetic moment during particle motion is of fundamental importance to the dynamics of magnetized plasma. The related rate of pitch angle scattering is investigated here for fast particles that thermally stream through static magnetic perturbations. For a uniform magnetic field with a localized perturbation it is found that the curvature parameter $\kappa^2=\min(R_c/\rho_L)$ does not predict the level of pitch angle scattering. Instead, based on numerical integration of particle orbits in prescribed magnetic fields, we derive predictions for the particle scattering rates, which can be characterized by the relative perturbation amplitude $A\simeq \delta B/B$, and the particle Larmor radius normalized by the field-aligned wavelength of the perturbations, $\rho_L/\lambda_{\|}$. Particles with 
$\rho_L/\lambda_{\|}\simeq1$, are subject to strong pitch angle scattering, while the level of scattering vanishes for both the limits of $\rho_L/\lambda_{\|} \ll 1$ and $\rho_L/\lambda_{\|} \gg 1$. The results are summarized in terms of a scattering operator, suitable for including the described scattering in basic kinetic models. 

\end{abstract}

\maketitle

\begin{quotation}

\end{quotation}


Fundamental properties of magnetized collisionless plasma are directly linked to the magnetic moment, $\mu =  m v_{\perp}^2/2B$, being an adiabatic invariant of the particle motion. By way of examples, the confinement times of particles in the Earth's radiation belts \cite{claudepierre:2020} as well plasmas in magnetic mirror machines \cite{post:1987} are governed directly by the time scales at which $\mu$ for the individual particles can be assumed to be constant.
During compression and rarefaction of solar wind plasma, dynamics related to the adiabatic invariance of $\mu$ is the cause of pressure anisotropy \cite{Chew:1956}. This anisotropy can drive both the mirror and firehose instabilities, which, in turn, can scatter the particle magnetic moments regulating the anisotropy towards the boundaries of marginal instability \cite{bale:2009}. At the Earth's bow shock {\sl in situ} spacecraft observations demonstrate directly how strong magnetic perturbations are the driver of electron pressure anisotropy and how bursts of Whistler waves can break the invariance of $\mu$ leading to pitch angle scattering, which in turn moderates the anisotropy \cite{oka:2017}. 

Generally for wave-particle interactions, the particles with orbit motions that by some measure are in phase with the wave dynamics will interact most strongly with the wave electric and magnetic fields and pitch angle scatter most efficiently. In contrast to wave-particle interaction, the adiabatic invariance of $\mu$ may also be destroyed during orbit motion in static magnetic configurations. The renowned study by Buchner 
\cite{buchner:1989} considers a static current sheet geometry relevant to the Earth's magnetotail and introduces the adiabatic parameter $\kappa^2=\min(R_c/\rho_L)$. Here $R_c$ is the radius of curvature for the magnetic field and $\rho_L= m\vp/(qB)$ is the particle gyro (Larmor) radius. For a given particle orbit, the value of $\kappa^2$ is evaluated at the location where $R_c/\rho_L$ is at its minimum, and the magnetic moment is generally well conserved for $\kappa^2>25$. For $10<\kappa^2< 25$, moderate changes in $\mu$ are typically observed,  whereas $\kappa^2<10$ is associated with chaotic particle motion and complete loss of $\mu$ as an adiabatic invariant for both electrons and ions. For magnetic reconnection, $\kappa^2$ has proven effective for understanding the electron dynamics, \cite{lavraud:2016} yielding a range of regimes for the structure of the electron diffusion region \cite{le:2013}. 

The present study is motivated in part by results for magnetic pumping \cite{barnes:1966}. In particular, it has recently been shown that strong Alfv\'enic magnetic perturbations $\delta B/B\simeq 1$ at the Earth's bow shock \cite{johlander:2017} have the ability to trap electrons, rendering magnetic pumping effective for heating superthermal electrons \cite{lichko:2020}. As is the case also for other Fermi heating models \cite{drake:2006,montag:2017}, the electron energization here requires a steady level of pitch angle scattering of superthermal electrons \cite{lichko:2020,egedal:2021} beyond scattering related to resonant whistler wave interactions \cite{oka:2017}. 
Similar to the analysis by Buchner\cite{buchner:1989} we will consider static magnetic configurations. In our configurations of magnetic perturbations, however, the overall direction of the magnetic field does not change and likely for this reason the numerical results presented are not well described by the $\kappa^2$-parameter. 

\begin{figure}[h!]
\includegraphics[width=0.47\textwidth]{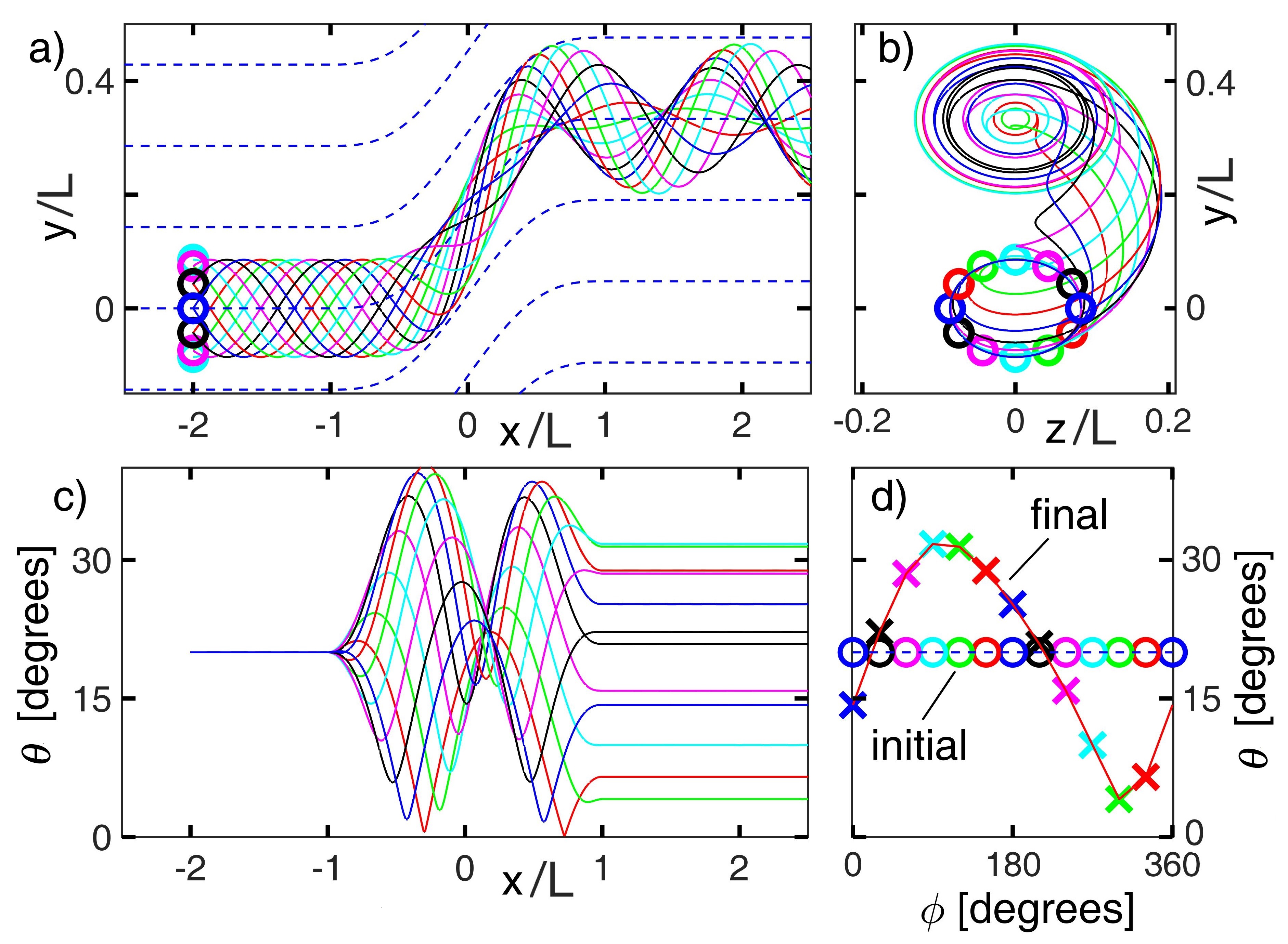}
\caption{Illustration of pitch angle diffusion for a selection of electrons, launched with identical initial guiding center parameters, but variable initial gyrophase. a,b) illustrate the particle orbits for the magnetic field given in Eq.~\eq{Bfield}. c,d) the evolution of the pitch angle $\theta$ as a function of $x/L$ and the initial azimuthal angle $\phi$ as the particles traverse the magnetic perturbation.}  
\label{fig:wiggle}
\end{figure}

We first explore the magnetic configurations as outlined by the dashed lines in Fig.~\ref{fig:wiggle}(a). Here a straight and uniform magnetic field in the $x$-direction is perturbed by a single localized and sinusoidal perpendicular pulse in the $y$-direction. More specifically, the magnetic field is given by the form
\begin{eqnarray}
\label{eq:Bfield}
  {\bf B}= \left\{\begin{array}{ccc}  B_0 \left[ 1,\, 0,\, 0\right]  & \text{for} &   |x|/L \geq 1 \\[2ex]
     B_0 \left[ 1,\, AF(x/L),\, 0\right]  & \text{for} &   |x|/L \leq 1 \end{array}
     \right.\quad,
\end{eqnarray}
where $F(x/L)= (\cos(\pi x/L)+1)/2$ provides the functional form of the perturbation applied in the $y$ direction with dimensionless amplitude, $A$. 
As observed in  Fig.~\ref{fig:wiggle}(a), across the interval $-L\leq x\leq L$ the field lines are offset in the $y$-direction by the distance $\Delta y = A\int_{-L}^L F(x) dx = AL$.  
In this configuration, numerical particle trajectories are initialized with  identical initial speeds  and 
pitch angles $\theta_i=\angle({\bf B},{\bf v}_i)=\arccos({\bf B}\cdot{\bf v}_i/Bv_i)=20^{\circ}$, but variable  azimuthal angles $\phi$ with respected to the $x$-direction.  The initial spatial location with $x/L=-2$ are in Figs.~\ref{fig:wiggle}(a,b)  arranged on a circle of radius $\rho_L/L = 0.1$ in the $yz$-plane,  such that all the particles share the same initial guiding center location. In Figs.~\ref{fig:wiggle}(a,b) it is apparent how the Larmor motion is centered on the initial field line. Particularly visible in Fig.~\ref{fig:wiggle}(b), the circular motions centered on $(y,z)=(0.3,0)$ after the particles transit the perturbation, are characterized by  variable radii and are indicative of pitch angle scattering. Note that because static magnetic fields do no work, the speeds of the particles are conserved.

In Figs.~\ref{fig:wiggle}(c,d) the scattering is analyzed in more detail, where in c) the lines represent the values of 
 $\theta=\arccos({\bf B}\cdot{\bf v}/Bv)$ recorded along the trajectories. Here, again,  all trajectories are initialized with $\theta_i=20^{\circ}$, but settle on a relative large ranges of values after the perturbation. In panel d) the final and initial values of $\theta$ are shown as a function of the initial azimuthal angle $\phi$. We observe how the final values of $\theta$ vary roughly $12^{\circ}$ about its initial value, corresponding to relatively strong scattering from this single perturbation.

To relate the above observations to the adiabatic parameter $\kappa^2$, the magnetic curvature of the configuration must be evaluated. In our analysis we consider relatively small values of $A<0.3$, and with $\bb=\B/B$ in this limit it follows that
\[
\bb\cdot\nabla \bb \simeq \fd{1}{B_0} \dd{}{x} \B = \left[0,\, -A\fd{\pi}{2L}\sin\left(\frac{\pi x}{L}\right),\,0\right] \quad.
\]
The local value for the radius of field-line curvature is $R_c=1/|\bb\cdot\nabla \bb|$, and its minimum value along the perturbation is $\min(R_c)= 2L/(A\pi)$, such that the adiabatic parameter becomes 
\begin{equation}
\label{eq:kappa}
\kappa^2 \simeq \fd{2}{A\pi} \fd{L}{\rho_L}\quad.
\end{equation}
For the orbits in Fig.~\ref{fig:wiggle} we use $A=0.3$ and $\rho_L/L=0.1$ such that $\kappa^2=21.2$ is relatively large. Thus, for the present configuration the $\kappa^2$ parameter predicts that the particle magnetic moments should be reasonably conserved, whereas the numerical results indicate that $\Delta\mu\simeq \mu$.

For a more general investigation, we numerically follow trajectories similar to those in Fig.~\ref{fig:wiggle}, but for a much larger range of initial conditions. As before, all trajectories are initialized at $x=-2L$, and we used the two dimensionless parameters $\rho_v/L = m v/(qB_0L)$, and $A$, as well as the two angles $\theta_i$ and $\phi$ to span the full range of orbits of the magnetic configuration.  Note that $\rho_v/L$ represents the particle speed normalized by a  characteristic speed  of the problem, $qB_0L/m$. We then characterize the spread in the change of pitch angles through
\begin{equation}
\label{eq:spread}
\sigma_1= \sqrt{\avep{(\theta-\theta_i)^2}}\quad,
\end{equation}
where for each set of parameters $(A, \rho_v/L,\theta_i)$ the $\phi$-average is over the results of 50 evenly spaced initial azimuthal angles. The subscript ``1'' on $\sigma_1$ denotes the results obtained from a single perturbation pulse, and will later be generalized to $\sigma_n$ for a train of $n$ perturbations.

\begin{figure}
     \centering
     \includegraphics[width=0.47\textwidth]{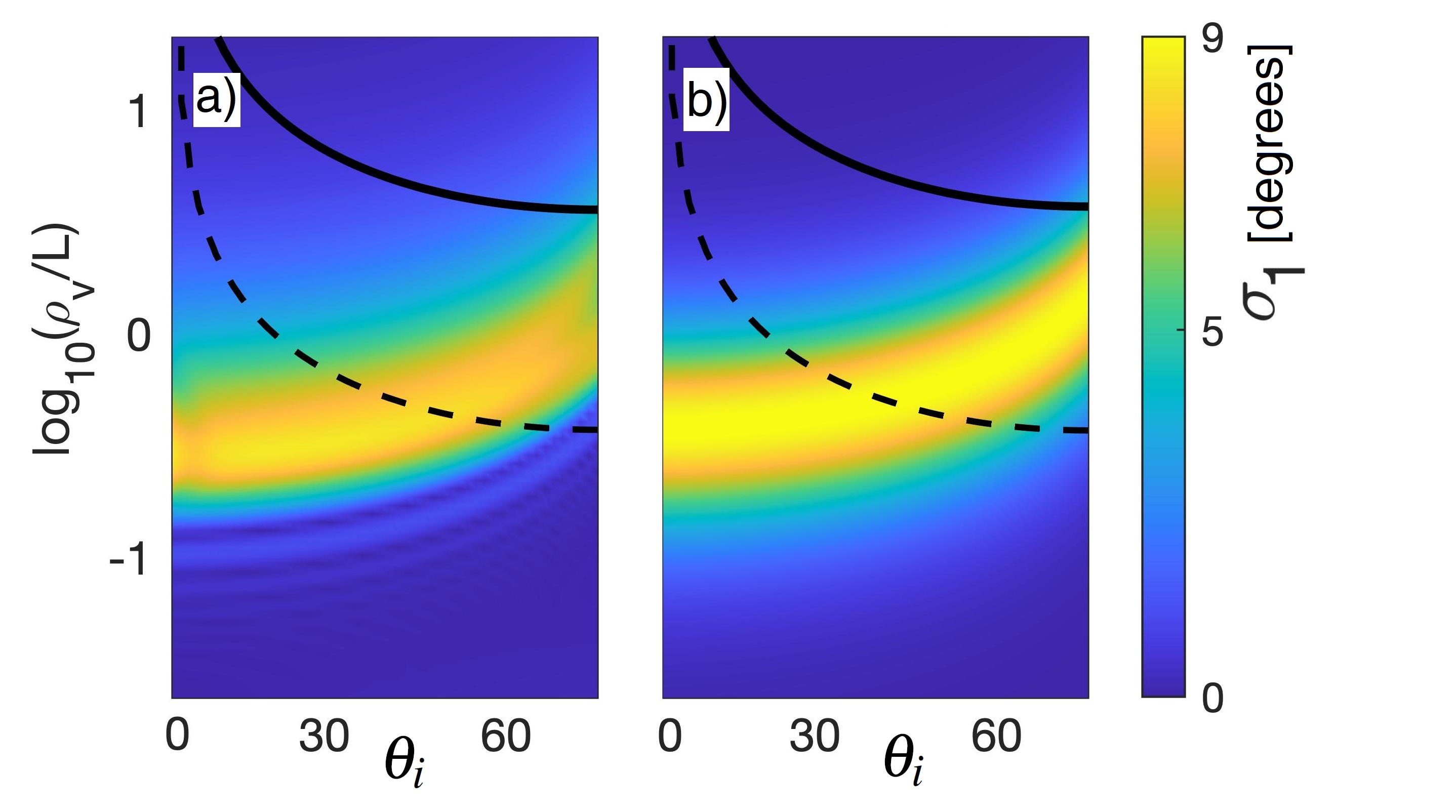}
             \caption{a) Color contours of $\sigma_1$ computed for $A=0.13$ as a function of the initial pitch angle $\theta_i$ and the parameter $\rho_v/L = mv/(qB_0L)$. Here $\sigma_1$ is defined in Eq.~\eq{spread}, and provides a measure of the spread in pitch angles observed by numerically integration of particle orbits. The black full and dashed lines correspond to $\kappa^2=1$ and $\kappa^2=10$, respectively, where $\kappa^2$ is the curvature parameter given in Eq.~\eq{kappa}. b) Color contours of approximate form for $\sigma_1$ introduced in Eq.~\eq{spreadapx}.}
              \label{fig:sigma1}
\end{figure}

The color contours in Fig.~\ref{fig:sigma1}(a) show the numerical results for $\sigma_1$ as a function of $(\theta_i, \log_{10}(\rho_v/L))$ obtained in a configuration described by Eq.~\eq{Bfield} for $A=0.13$. The solid black line indicates values of $(\theta_i, \log_{10}(\rho_v/L))$ for which $\kappa^2=1$, whereas $\kappa^2=10$ at the locations of the the dashed black line. It is observed that these contour lines of $\kappa^2$ do not correlate well with the location of the ribbon where $\sigma_1\simeq 9$. 
Rather, the upward trend of the ribbon as a function of increasing $\theta_i$ indicates that the scattering level is a function of $A$ and $\rho_{\|}/L \equiv m \vz/(qB_0L)$, and not $\rho_L/L$ as it appears in Eq.~\eq{kappa}.

Through curve fitting we identify an approximation for the scattering rate by the simple form 
\begin{equation}
\label{eq:spreadapx}
\sigma_1\simeq 280^{\circ} A \left(\fd{2\rho_{\|}/L}{1 + ( 2\rho_{\|}/L)^2}\right)^2\quad.
\end{equation}
In contrast to $\kappa^2$ that predicts the strongest scattering for the largest values of $\rho_L/L$, as seen in Fig.~\ref{fig:sigma1}(a) for the considered geometry, and captured by the expression for $\sigma_1$ illustrated in Fig.~\ref{fig:sigma1}(b), the strongest pitch angle scattering is observed when $\rho_{\|}\simeq L$. More physically, the pitch angle scattering by the localized perturbations is most effective for particles which have ``parallel Larmor'' radii similar to the perturbation length, and from here the scattering efficiency decreases rapidly with increasing values of $\vz$. The latter is observed by considering cuts in Fig.~\ref{fig:sigma1} for fixed value of $\theta_i$, such that $\vz\propto \rho_{\|}/L$. Again, this difference compared to the expectations based on $\kappa^2$ is most likely related to the localized nature of our perturbation, whereas $\kappa^2$ is derived for configurations where the directions of $\B$ changes at the global scale. 

\begin{figure}
     \centering
     \includegraphics[width=0.47\textwidth]{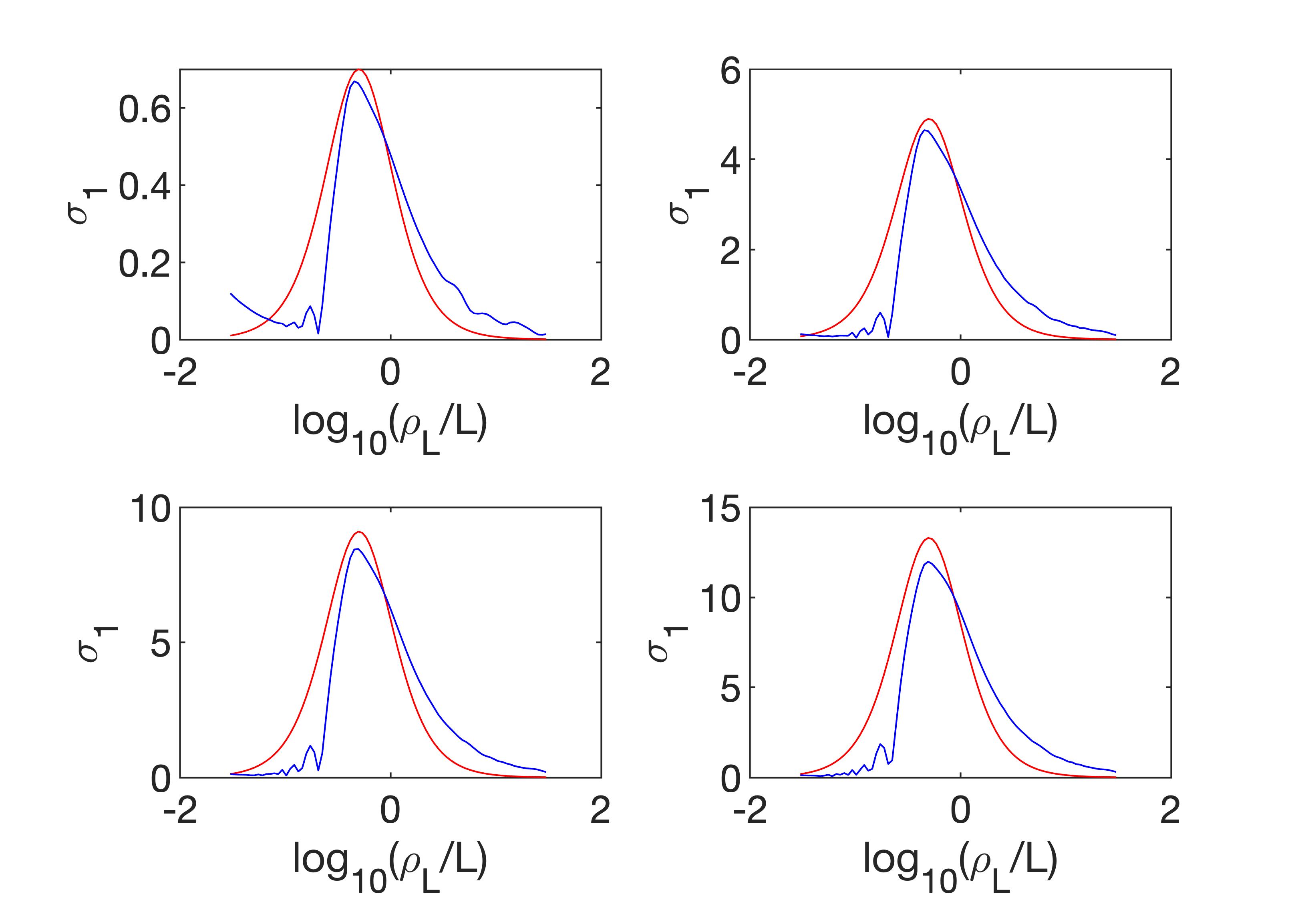}
         \caption{Blue lines represent $\sigma_1$ defined in Eq.~\eq{spread} as a function of $\rho_v/L$, obtained through numerical integration of orbits in the magnetic field of Eq.~\eq{Bfield} for $\theta_i=40^{\circ}$ and amplitudes $A\in\{ 0.01, 0.07, 0.13, 0.19\}$. The similar red lines are obtained from the approximate form in Eq.~\eq{spreadapx}.}
 \label{fig:cut2}
\end{figure}

Numerically, for other values of $A$ we obtain results similar to those in Fig.~\ref{fig:sigma1}(a), with the values of $\sigma_1$ scaling linearly with $A$. Thus, Eq.~\eq{spreadapx} is appropriate not only for $A=0.13$ (as considered in Fig.~\ref{fig:sigma1}), but remains a good approximation for the range of $0\leq A\leq 0.4$.
This is demonstrated in part in Fig.~\ref{fig:cut2} where both the numerical (blue lines) and functional form (red lines) for $\sigma_1$ are evaluated as a function of $\log_{10}(\rho_{\|}/L)$ for $\theta_i=40^\circ$ and $A\in \{ 0.01, 0.07, 0.13, 0.19\}$.

For a perturbations in terms of a standard sinusoidal wave at a fixed wave-number, the scattering by the various ``$B_y$-pulses'' can be expected to be correlated and may reduce the level of scattering. However, for most systems of interest magnetic perturbations are not monochromatic such that the scattering by the individual half-periods of the perturbations will become uncorrelated. The result can then be characterized as a random walk where after $n$ steps the statistical spread is expect to be 
\begin{equation}
    \label{eq:nspread}
    \sigma_n= \sqrt{n}\,\sigma_1\quad.
\end{equation}

\begin{figure}[h!]
\centering
\includegraphics[scale=0.36]{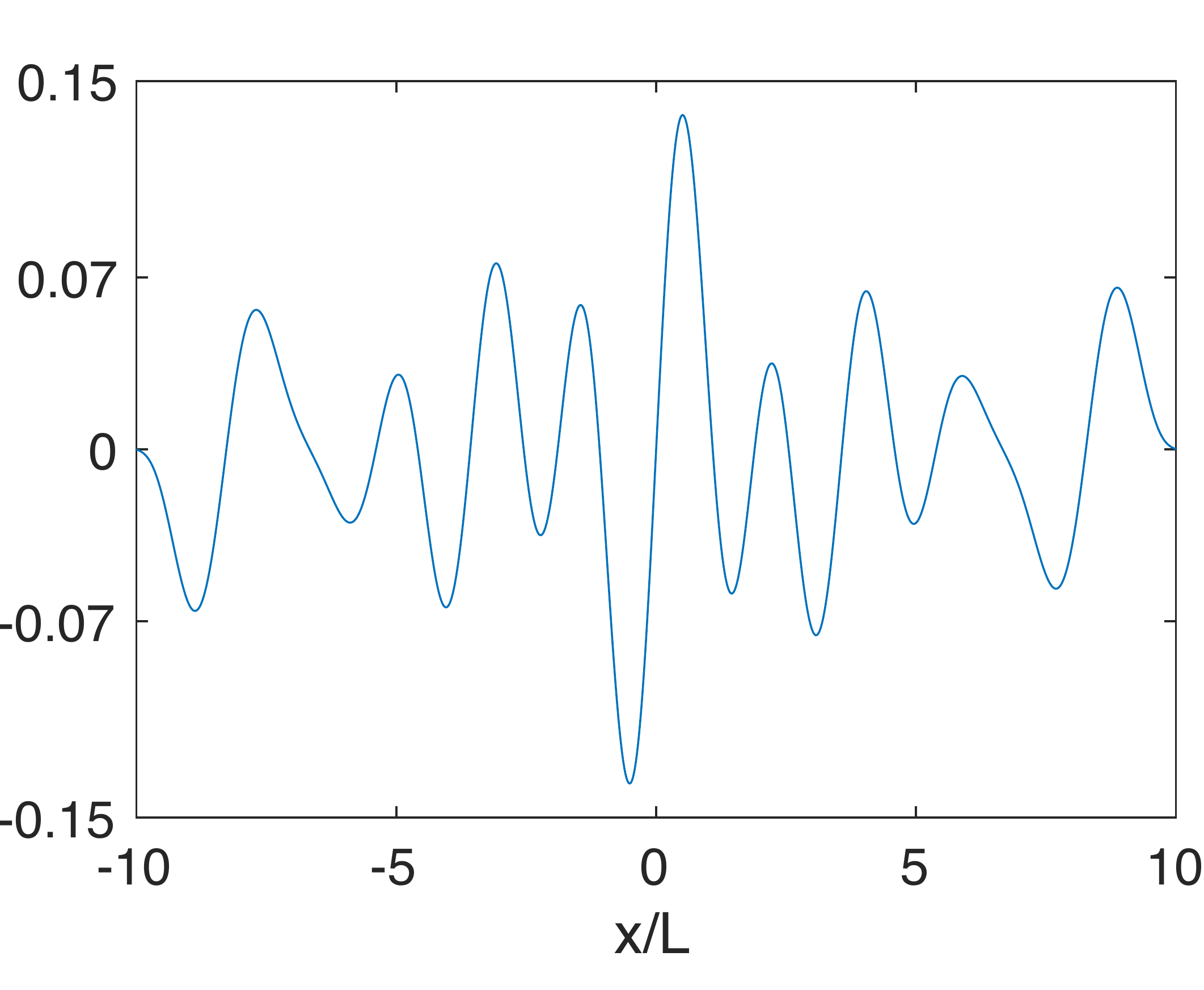}
\caption{Perturbation applied for the $B_y/B_0 =F(x/L)$, as specified in Eqs.~\eq{Bfield2} and \eq{perturb}.}
\label{fig:wiggle_2}
\end{figure}

To test the applicability of Eq.~\eq{nspread} we consider the more complicated magnetic geometry given by 
\begin{eqnarray}
\label{eq:Bfield2}
  {\bf B}= \left\{\begin{array}{ccc}  B_0 \left[ 1,\, 0,\, 0\right]  & \text{for} &   |x|/L \geq 10 \\[2ex]
     B_0 \left[ 1,\, F(x/L),\, 0\right]  & \text{for} &   |x|/L \leq 10 \end{array}
     \right.\quad.
\end{eqnarray}
 with $F(x/L)$ being a sum of multiple sinusoidal perturbations with variable wave lengths. More specifically, with $\tx=\pi x/L$, we use 
\begin{eqnarray}
\label{eq:perturb}
F(x/L)&=&0.03\left(\sin\,0.5 \tx\,+\,\sin\,0.7\tx \,+\, \sin\, \tx\right.
\nonumber \\[1ex]
& & \quad\left. \,+\,\sin\, 1.1\tx\,+\,\sin\,1.2\tx\right)\quad.
\end{eqnarray}
This form is illustrated in Fig.~\ref{fig:wiggle_2}, and  includes 18 local maxima and minima, characterized by an average absolute magnitude of about $A=0.05$. 

\begin{figure}
         \centering
       \includegraphics[width=0.45\textwidth]{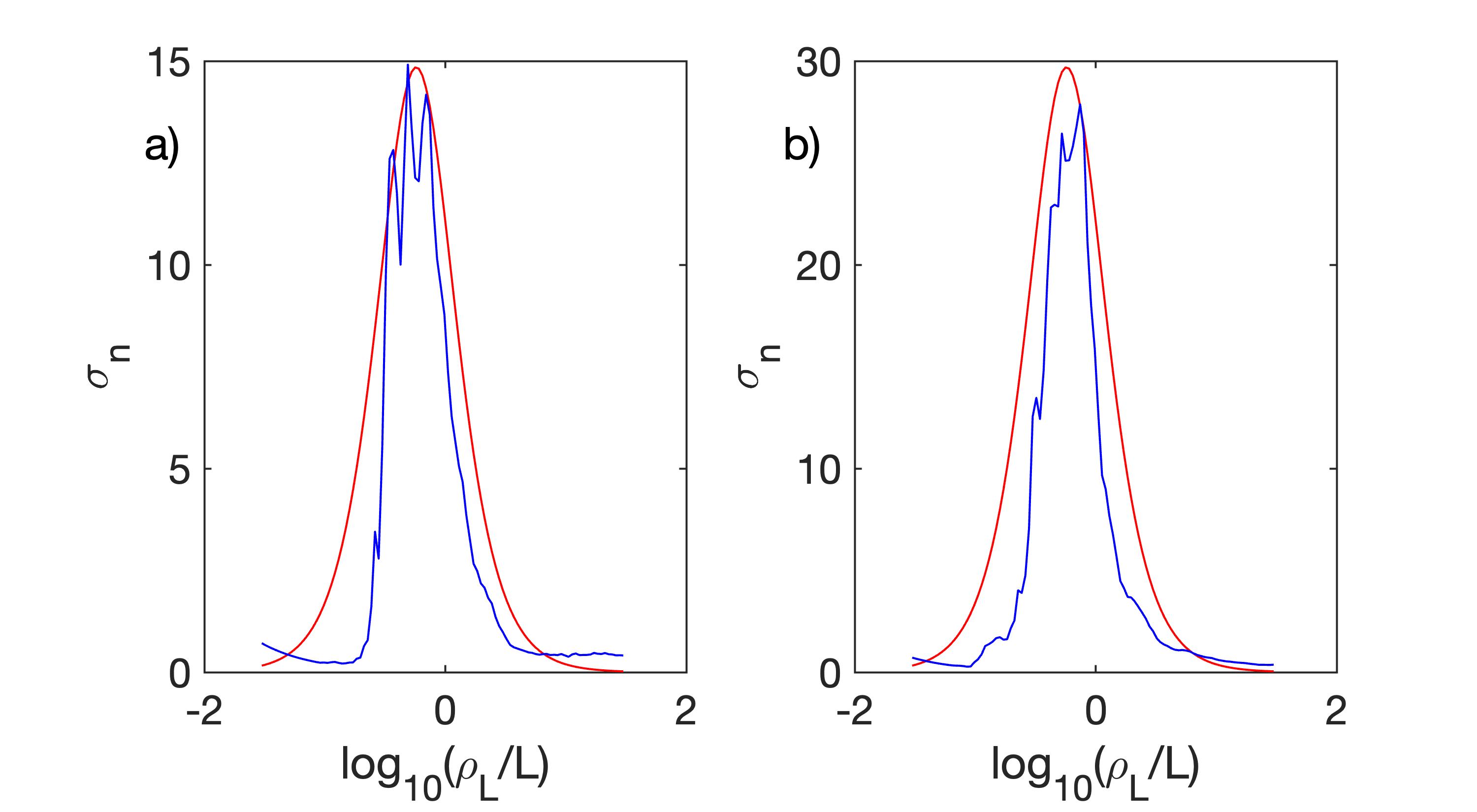}
         \caption{a) The blue line show $\left<\sigma\right>_{\theta_i}$ as defined in Eq.~\eq{spreadave} and numerically evaluated as  function of $\log_{10}(\rho_L/L)$  for the field configuration specified in Eqs.~\eq{Bfield2} and \eq{perturb}, and illustrated in Fig.~\ref{fig:wiggle_2}. The red line is the prediction by $\sigma_n$ in Eq.~\eq{nspread}, evaluated with $n=18$ and $A=0.05$. b) Similar to a) but obtained with the perturbation $F(x/L)$ in Eq.~\eq{perturb} doubled.  }
         \label{fig:wiggle_3}
\end{figure}

Considering the configuration described by Eqs.~\eq{Bfield2} and \eq{perturb}, the blue line in Fig.~\ref{fig:wiggle_3}(a) represents the numerical result for 
\begin{equation}
\label{eq:spreadave}
\left<\sigma\right>_{\theta_i} = 
\left<\sqrt{\avep{(\theta-\theta_i)^2}}\right>_{\theta_i} 
\quad,
\end{equation}
where compared to Eq.~\eq{spread} we take the additional average over $\theta_i$ for the range of 
$0\leq \theta_i\leq 70^{\circ}$. Based on the results above for a single perturbation pulse, we expect that particles will here scatter with an efficiency corresponding to $\sigma_{18}=\sqrt{18}\sigma_1$, with $\sigma_1$ given in  Eq.~\eq{spreadapx} and evaluated with $A=0.05$. This theoretical estimate yields the red curve in Fig.~\ref{fig:wiggle_3}(a), providing a reasonable approximation for the numerical results.  The similar results in 
Fig.~\ref{fig:wiggle_3}(b) are obtained for the same functional form of $F(x/L)$ in Eq.~\eq{perturb}, but with its magnitude increased by a factor of two.

Based on the above results we can now construct a scattering operator which we envision will be useful in crude kinetic calculations, and for order of magnitude estimates for the effectiveness of this type of scattering in naturally occurring systems. For a gyrotropic distribution of particles characterized by $f(v,\theta)$  by Fick's law the flux of particles in velocity space induced by ``$\theta$-diffusion'' is given by 
\[  {\bf J}_v=-D\frac{1}{v}\frac{\partial}{\partial \theta}f\hat{\theta}\,\,,\quad D=\frac{(v\Delta\theta)^2}{\tau}\quad,\]
applicable to a system with average random  step sizes, $v\Delta\theta$, expected for the time scale $\tau$. Recasting the continuity condition $\ddt{f}{t}+ \nabla\cdot{\bf J}_v=0$ in terms of $\xi=\vz/v=\cos \theta$, simple manipulations then yield
\[
\dd{f}{t} = \fd{\Delta \theta^2}{\tau} {\cal L} f\,,\quad
{\cal L} = \dd{}{\xi}(1-\xi^2)\dd{}{\xi}\quad,
\]
where ${\cal L}$ is the well-known Lorentz scattering operator. For a system characterized by magnetic perturbations of relative size  $\Delta B/B=A$, around a wave number $k (= 2\pi/\lambda)$, the particle thermal transit time through half a wavelength will be on the order of $\tau= \lambda/(2v_{\|})\simeq \lambda/(2\rho_v\omega_c)=\pi/( k\rho_v\omega_c)$, where $\omega_c= qB/m$ is the gyro-frequency. The expected step size in $\theta$ is $\Delta \theta = (\pi/180^{\circ}) \sigma_1$, with $\sigma_1$ in Eq.~\eq{spread} evaluated with $\rho_{\|}/L \simeq k\rho_v/\pi$. With these approximations the wave induced scattering can then be characterized through the operator
\begin{equation}
    \label{eq:C}
     C = \fd{\Delta \theta^2}{\tau}  {\cal L} \simeq 37\, \omega_c \,A^2 \fd{(2k\rho_v/\pi)^5}{(1+ (2k\rho_v/\pi)^2)^4 }\, {\cal L}\quad,
\end{equation}
where again $\rho_v = mv/qB$.  The approximation for $ C$ is expected to be valid for $0.03 \lesssim k \rho_v \lesssim 30$, and beyond this range the use of the form should be validated through additional numerical calculations.

In summary, we have numerically investigated the scattering in pitch angle of particles traversing a simple localized magnetic perturbation imposed on a large scale uniform magnetic field. For this configuration the curvature parameter, $\kappa^2$, does not provide useful information about the scattering levels. Rather, we find that the scattering is most effective for particles with Larmor radii similar to the magnetic field aligned scale size of the perturbation, $\rho_L\simeq \lambda_{\|}$. A simple analytic form is introduced to describe the scattering level of a single magnetic pulse perturbation, and through random walk arguments, the form generalizes to more physical systems including multiple waves at variable amplitudes and wave length. Finally, the numerical results are summarized in an analytical form of a pitch angle scattering operator by which this type of scattering can be included wider kinetic description of the plasma dynamics.

\begin{acknowledgments}
This work was supported in part by
a H.~I.~Romnes Faculty Fellowship by the UW-Madison Office of the Vice
    Chancellor for Research and Graduate Education, and the NASA HERMES DRIVE Science Center grant No. 80NSSC20K0604. 
\end{acknowledgments}

\section*{Conflicts of interest}
    The authors have no conflicts to disclose.

\bibliographystyle{unsrt}   
\bibliography{references}

\end{document}